\documentclass[slac_one]{revtex4}
\usepackage{graphicx}
\usepackage{fancyhdr}
\begin{document}

\title{Top Quark Studies with the first CMS Data} 

%

\author{Frank-Peter Schilling (on behalf of the CMS Collaboration)}
\affiliation{Karlsruhe Institute of Technology (KIT), Karlsruhe, GERMANY}

\begin{abstract}

Studies are presented of the selection of events consistent with top
quark pair production in data recorded by the CMS detector at the LHC, corresponding to an integrated luminosity of $0.84\pm 0.09 \rm\ pb^{-1}$
and at center-of-mass energy $\sqrt{s}= 7 \rm\ TeV$.  Results
are presented for the lepton+jets as well as dilepton channels. 
Event yields in data are compared to those in simulation, and several background processes are estimated using data-driven techniques. The observed yields of top-antitop candidate events are roughly consistent with the Standard Model.

\end{abstract}

\maketitle

\thispagestyle{fancy}



\section{INTRODUCTION}

Top quark physics is an important part of the research program at the LHC. 
Since its discovery in proton-antiproton collisions at the Tevatron collider, the properties of the top quark have been
studied in detail (see e.g. \cite{topreview} and references therein).
With the advent of the Large Hadron Collider (LHC), top-quark processes can be studied for the first time in multi-TeV collisions.  

Due to its large mass~\cite{topmass}, the top quark may play a special role in the standard model.
The top quark decays rapidly, long
before having the chance to form a bound state hadron.  Hence, it allows direct access to measurements of its
mass, spin, charge and other properties.  Additionally, since the
Higgs boson of the standard model couples to fermions in strength
proportional to the fermion's mass, the Higgs coupling to the top
quark is large.  Because of this, detailed study of the properties of
the top quark can provide constraints on the yet-to-be observed
Higgs boson.  Further, in light of its large mass, it is
hypothesized that the top quark could play a role in electroweak
symmetry breaking and the generation of particle masses in
alternatives to the Higgs mechanism.  Finally, several signatures of
new physics accessible at the LHC either suffer from top-quark
production as a significant background or contain top quarks
themselves.

At the LHC, the top quark is expected to be produced primarily via the
strong interaction (mostly via gluon-gluon fusion, in contrast to the Tevatron) in ${\rm t{\bar t}}$ pairs.  
The next-to-leading order (NLO) corrections to top-quark pair production at
hadron colliders were calculated for unobserved spins in~\cite{nason,beenakker} and
with the full top-quark spin dependence in~\cite{bernreuther1,bernreuther2}. Only recently a
complete analytic result for the NLO partonic cross section has
been published~\cite{czakon}. Approximations towards a full NNLO result have
been obtained by various groups, e.g.~\cite{Cacciari:2008zb,Kidonakis:2008mu,Moch:2008qy}.

Within the standard model, the top quark decays via the weak process
${\rm t\rightarrow Wb}$ nearly 100\% of the time.  Subsequently
top-quark pair events are categorized according to the decay of the
two ${\rm W}$ bosons. We consider here the dilepton channel, in which both W bosons decay to leptons, and the lepton+jets channel, where one W
 decays leptonically, while the other one decays into quarks.

In this note, the first results~\cite{top-10-004,toptwiki} on top quark physics obtained with the initial 7 TeV LHC data are presented.
Previous simulation studies for both channels can be found in~\cite{top09002,top09003,top09004}. 


\section{DATA AND SIMULATED SAMPLES}

The selected sample corresponds to an integrated luminosity~\cite{lumipas} of $0.84\pm0.09 \rm\ pb^{-1}$, using data recorded by CMS~\cite{cmsdet} up to August 2010. 
Before being used in the analysis, data events are constrained  to periods
in which the CMS detector was fully operational.
Additionally, events are vetoed if they are identified as resulting
from beam halo interactions or from beam scraping.  Finally, events
are required to possess at least one well-constructed primary vertex
within $|z|<$15~cm.  

Simulated samples of top-quark pair production events are made using the
MADGRAPH event generator~\cite{madgraph}, subsequently processed with
PYTHIA~\cite{pythia}, and then processed with a full CMS detector
simulation based on GEANT4~\cite{geant}.  Events are generated
with up to four additional hard partons.
Various background samples were produced. MADGRAPH is used for $W$/$Z$/$\gamma$+jets
production and single top.  Leptonic tau decays are included in the
Drell-Yan samples.  PYTHIA is used to generate QCD events used in the
study of the multijet backgrounds.

The top-quark pair production simulation has been normalized using a NLO cross section 
of 
$\sigma_{t\bar{t}}=157.5^{+23.2}_{-24.4} \rm\ pb$, obtained using
MCFM~\cite{mcfm,mcfm:tt}.  The uncertainty in the cross section includes the
scale uncertainties, determined by varying the factorization and
renormalization scales by a factor 2 and 0.5 around the central scale choice 
of $m_t=172.5 \rm\ GeV/c^2$, and the uncertainties from the parton distribution functions and the value of the strong coupling $\alpha_S$.
Similarly, the simulations of $W$/$Z$/$\gamma$+jets
production and single top production have been normalized using available inclusive N(N)LO cross section calculations.


\section{DILEPTON CHANNEL}

Dilepton events in the dielectron ($ee$), dimuon ($\mu\mu$) and electron-muon ($e\mu$) modes are considered. Events passing a single muon or electron  trigger are selected which contain two oppositely charged, high transverse momentum $p_T$ leptons with $p_T>20 \rm\ GeV/c$ and pseudorapidity $|\eta|<2.5 \, (2.4)$ for muons (electrons). Muons reconstructed ~\cite{muons} with high quality are selected, whereas identification based on cluster shape properties and track-cluster matching criteria is applied to electrons~\cite{electrons}, and electron candidates consistent with photon conversions are rejected. Leptons are required to be isolated within a cone of $\Delta R = \sqrt{\Delta\eta^2 + \Delta\phi^2} <0.3$, using a relative isolation variable which employs sums of track transverse momenta and calorimeter transverse energy deposits, scaled to the lepton $p_T$.
Both leptons are required to be consistent with originating from the primary hard interaction, both in the transverse plane as well as along the beam direction. For $ee$ and $\mu\mu$ candidates, the dilepton invariant mass is required to satisfy $|M_{ll}-M_{Z}|>15 \rm\ GeV/c^2$, to reject Z events. Missing transverse energy (MET)~\cite{met} is calculated from calorimeter signals, made more accurate by applying a track-based correction for the inexact calorimeter response. A cut ${\rm MET}>30 \, (20) \rm\ GeV$ is applied in the $ee,\mu\mu$ ($e\mu$) channels. Jets~\cite{jets} are clustered using the anti-$k_T$ algorithm~\cite{antikt} with $R=0.5$, using calorimeter information and corrected using tracker measurements. Jet energies are corrected to achieve uniform response in $\eta$ (relative) and $p_T$ (absolute). The jet energy scale uncertainty is estimated as $5\%$. Jets are required to satisfy $p_T>30 \rm\ GeV/c$ and $|\eta|<2.5$ and must not overlap with any electron or muon within $\Delta R<0.4$. At least two jets are requested for the full event selection.

\begin{figure}[tb]
\centering
\includegraphics[width=0.44\linewidth]{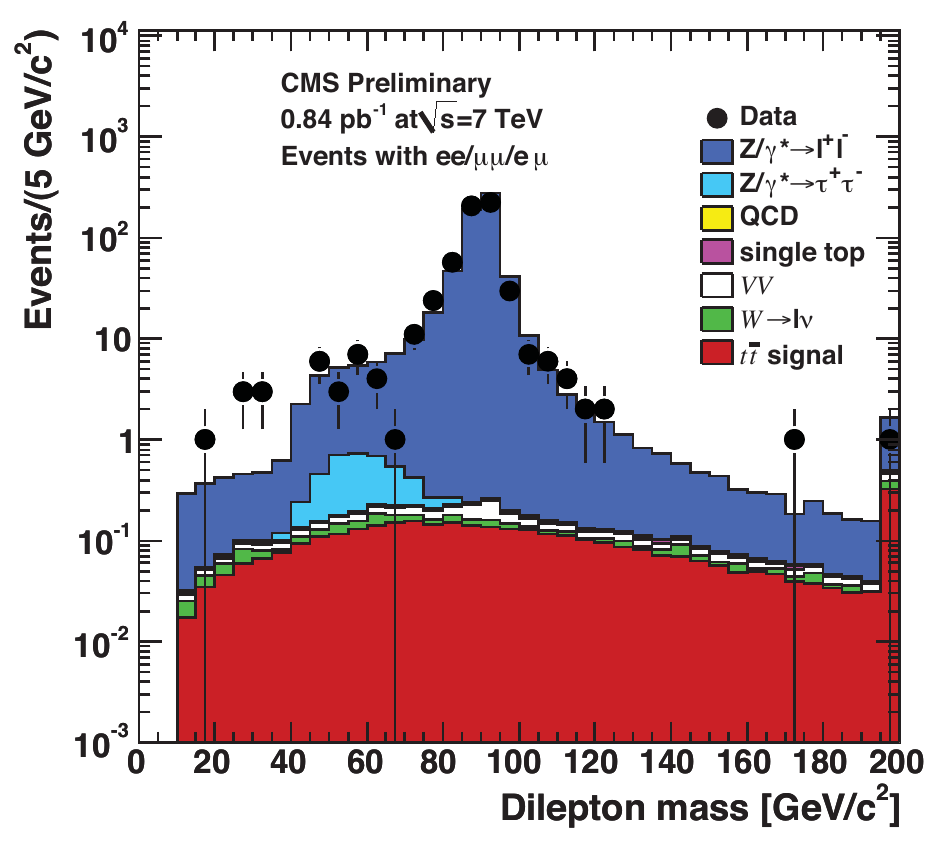}
\includegraphics[width=0.45\linewidth]{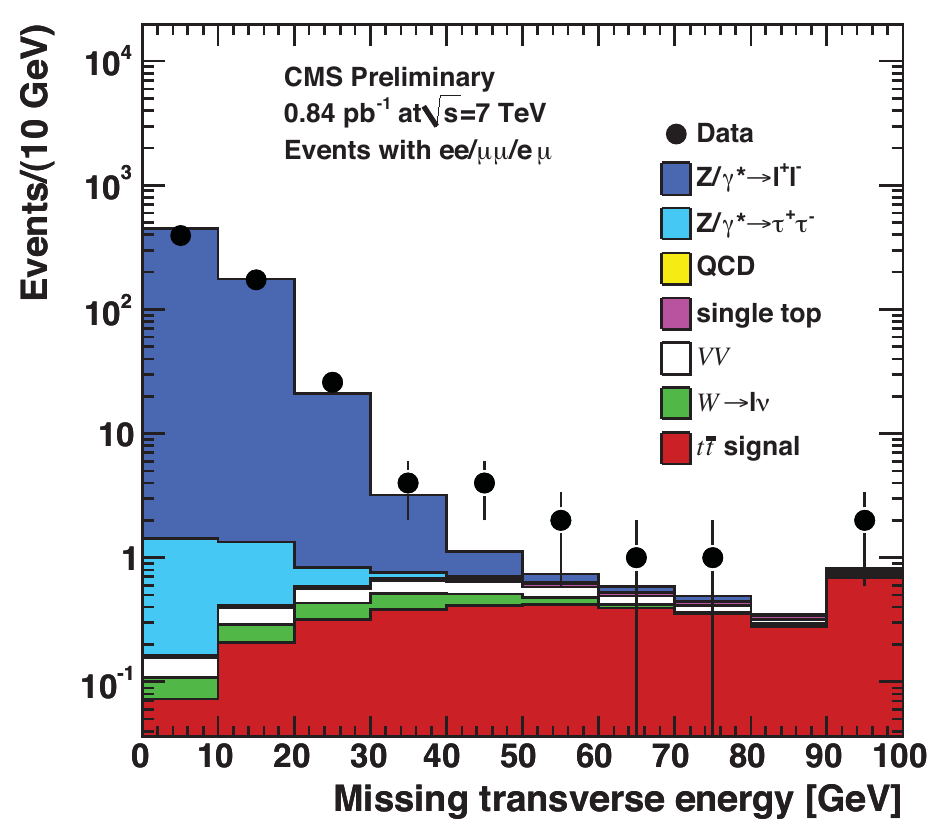}
\caption{Distributions of dilepton invariant mass (left) and missing transverse energy (right) for a relaxed event selection, as described in the text.}
\label{fig:dil-relaxed}
\end{figure}

Distributions for a relaxed event selection (without jets+MET requirements, no Z veto applied) are shown in Figure~\ref{fig:dil-relaxed}. Good agreement is observed between data and simulation, scaled to the integrated luminosity of the data.

For several sources of backgrounds, data-driven estimation techniques are tested. Drell-Yan events passing the Z veto are estimated by counting events rejected by this veto, scaled by the ratio of events outside and inside the veto region, obtained from simulation. The estimated systematic uncertainty of the method is $50\%$. 

Background from events with non-genuine isolated leptons (i.e. not originating from W/Z decays) is estimated by weighting events passing loose lepton identification with a tight-to-loose ratio which is parameterized in $p_T$ and $\eta$, measured in an inclusive QCD sample. The method is used to estimate the contributions from QCD multi-jet and W+jets events, containing two and one non-genuine lepton respectively, with a $50\%$ systematic uncertainty per lepton. The data-driven estimates are in reasonable agreement with expectations from simulation.

\begin{figure}[tb]
\centering
\includegraphics[width=0.45\linewidth]{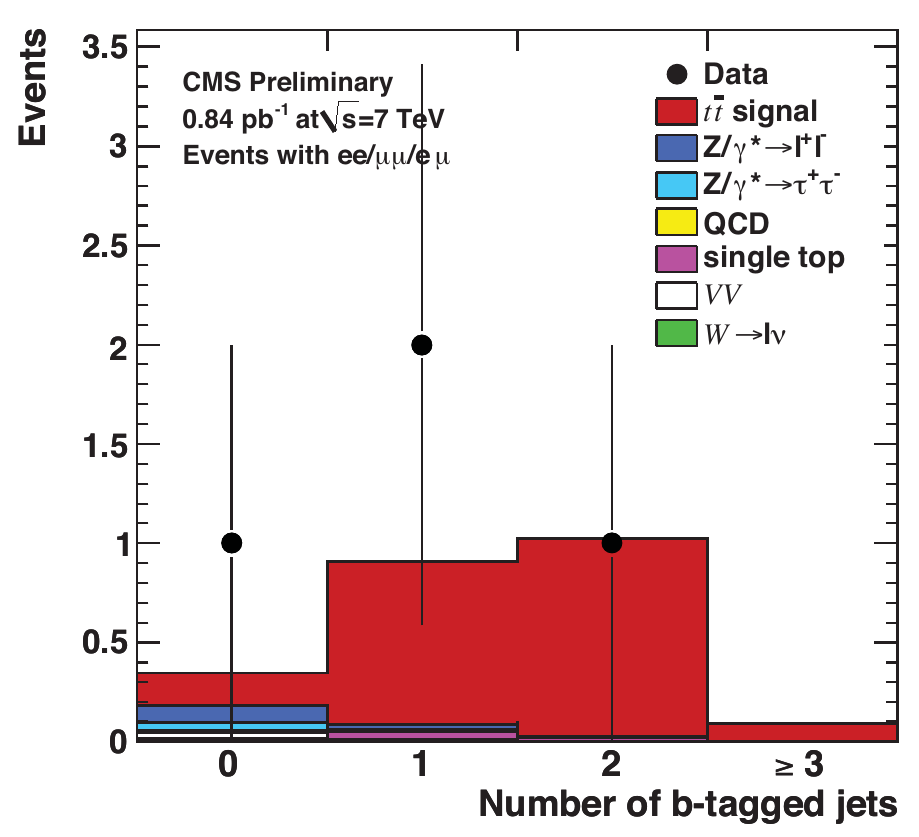}
\includegraphics[width=0.44\linewidth]{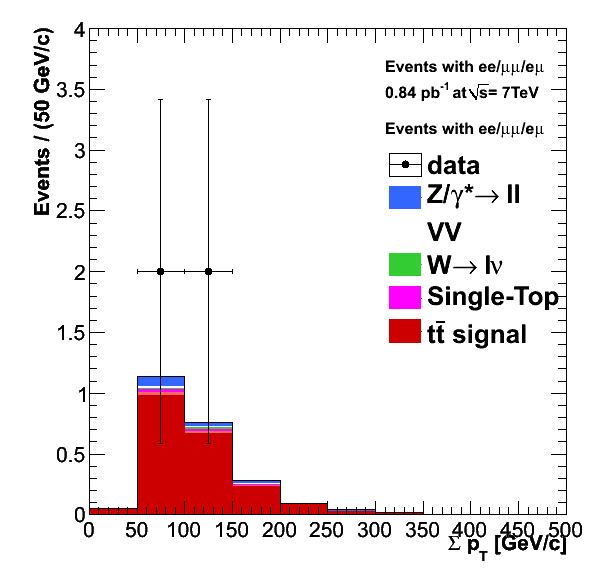}
\caption{Distributions of dilepton candidates passing the full event selection: number of b-tagged jets (left) and scalar lepton $p_T$ sum (right).}
\label{fig:dil-full}
\end{figure}

Applying the full event selection, including Z-veto, MET requirement and requesting at least two jets, there are four events selected in the sample.
The expected non-top background from simulation is less than 0.3 events, while 2.1 top signal events are expected.
Figure~\ref{fig:dil-full} shows the b-jet multiplicity distribution using a b-tagging algorithm~\cite{btag} based on the impact parameter significances of the tracks associated with the jets. A loose working point with $80\%$ b-jet efficiency and $10\%$ mistagging rate in QCD events is used. Also shown is the distribution of the scalar lepton $p_T$ sum. The observed events are consistent with a top-antitop hypothesis.

\section{LEPTON+JETS CHANNEL}

In the lepton+jets channel, both the $e$+jets and $\mu$+jets modes are considered. 
Events are selected which contain exactly one isolated, high-$p_T$ lepton. For $e$+jets, electrons passing tight identification criteria, inconsistent with originating from photon conversions and fulfilling $p_T>30 \rm\  GeV/c$ and $|\eta|<2.4$ are selected. For $\mu$+jets, high quality muons with 
$p_T>20 \rm\  GeV/c$ and $|\eta|<2.1$ are considered. Jets and MET are reconstructed using calorimeter information. Jets are required to have $p_T>30 \rm\ GeV/c$ and $|\eta|<2.4$. There is no explicit MET requirement. At least four jets are expected for signal.

\begin{figure}[tb]
\centering
\includegraphics[angle=90,width=0.46\linewidth]{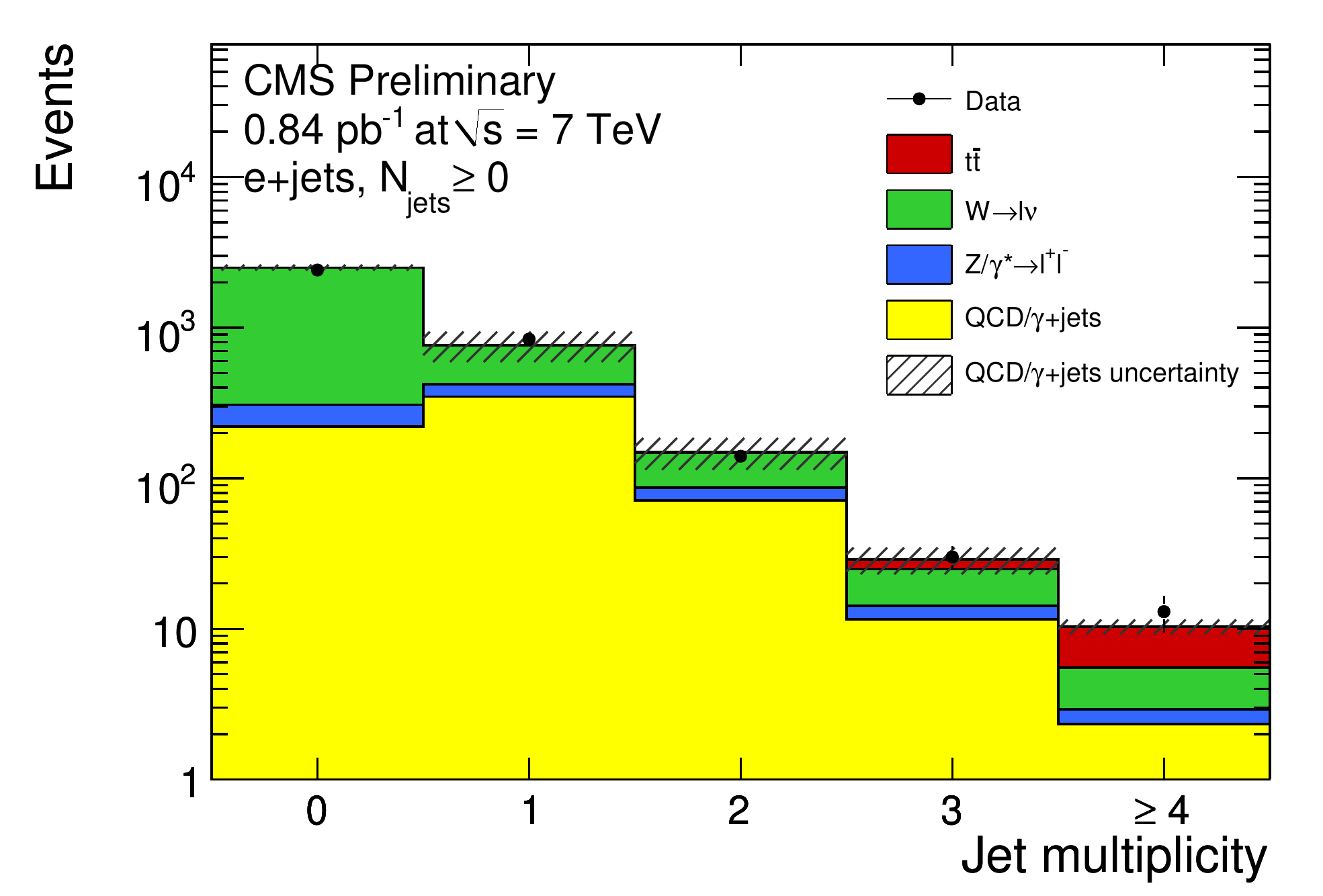}
\includegraphics[angle=90,width=0.46\linewidth]{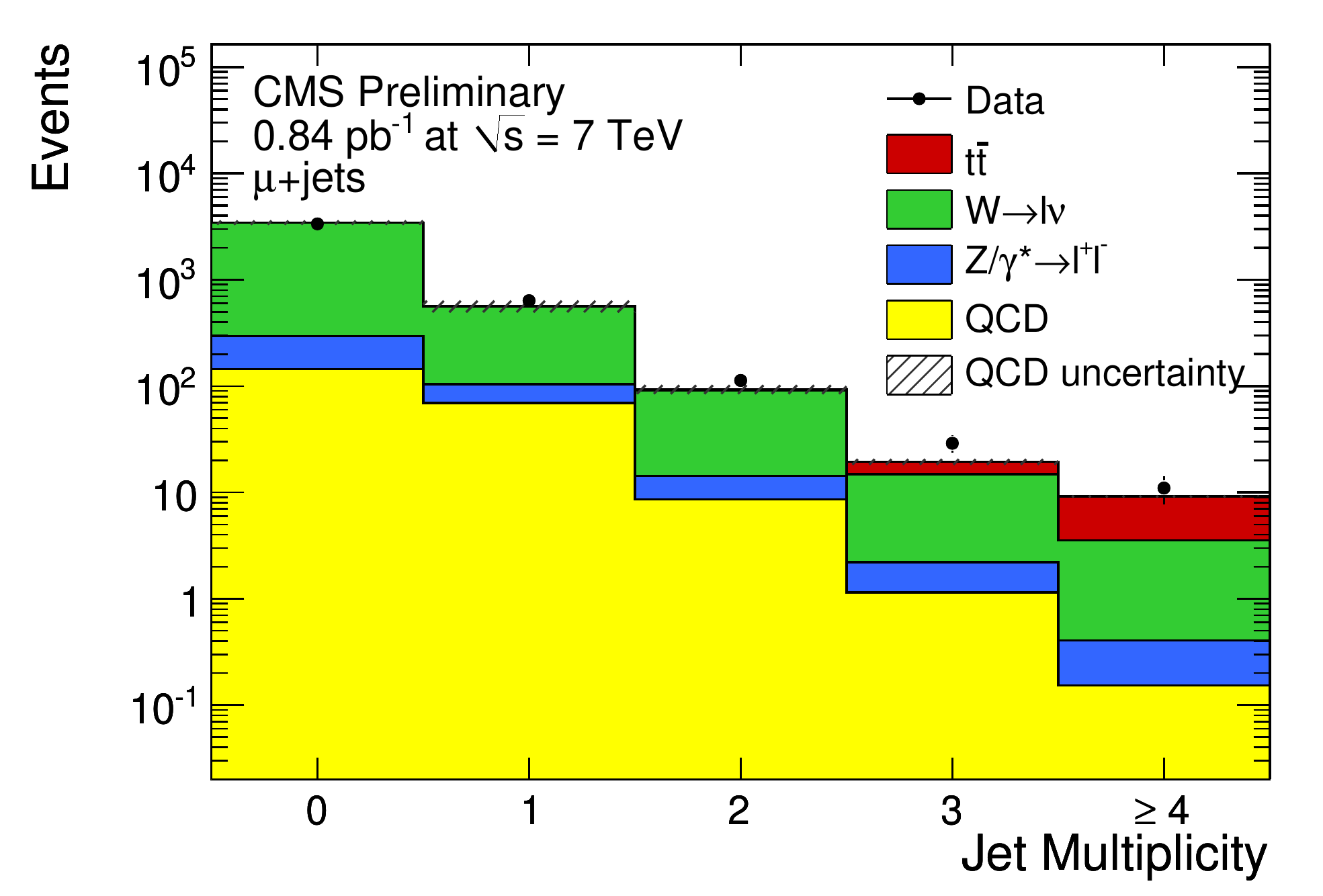}
\caption{Jet multiplicity for the e+jets (left) and $\mu$+jets (right) event selections, without b-tagging.}
\label{fig:ljets-njets}
\end{figure}

Event yields as a function of jet multiplicity are shown in Figure~\ref{fig:ljets-njets}. 
  Good agreement between data and simulation is observed in all jet bins.
The top signal-to-background ratio is increasingly significant at high jet multiplicities.

Differential distributions (e.g. muon transverse momentum $p_T^{\mu}$, MET, transverse W-mass; not shown) for different jet multiplicities  are also in agreement with the data, with the exception of the low $p_T^{\mu}$, low MET region at low $N_{\rm jets}$ in $\mu$+jets.
In this kinematic region, QCD multi-jet background is contributing, which is 
not expected to be realistically modeled by simulation. While there is a deficit of predicted QCD background in $\mu$+jets, data and simulation are in better agreement in $e$+jets. This may be explained by the different sources of background contributing in $\mu$+jets (muons from semi-leptonic $b$-decays and decays-in-flight) and $e$+jets (mostly photon conversions), respectively.

Several methods are studied which allow a data-driven estimate of the amount of QCD background in the selected sample. In both $e$+jets and $\mu$+jets, a method based on the relative isolation variable is employed. The isolation distribution is fitted with a suitable function in the non-isolated (QCD dominated) sideband region, which is then extrapolated into the isolated signal region. Another method, often referred to as {\em ABCD method}, exploits two nearly uncorrelated variables (here lepton impact parameter and relative isolation) which separate signal and QCD background in $\mu$+jets. A third method, applied in $e$+jets, is based on a template fit of the MET or 
$H_{T,lep}={\rm MET}+p_{T,lep}$ distribution, using a data-driven QCD template.
Two models are considered in order to obtain template
distributions for QCD multijet events: ``background'' electrons, in
which the electron candidate very nearly satisfies the selection
criteria but instead is a marginal failure; and jet-electrons,
positively identified jet objects with large electromagnetic fraction,
that closely resemble electron candidates. In $e$+jets ($\mu$+jets), a 50\% (100\%) systematic uncertainty is assigned to the QCD background, based on the data-driven estimates.

\begin{figure}[tb]
\centering
\includegraphics[angle=0,width=0.45\linewidth]{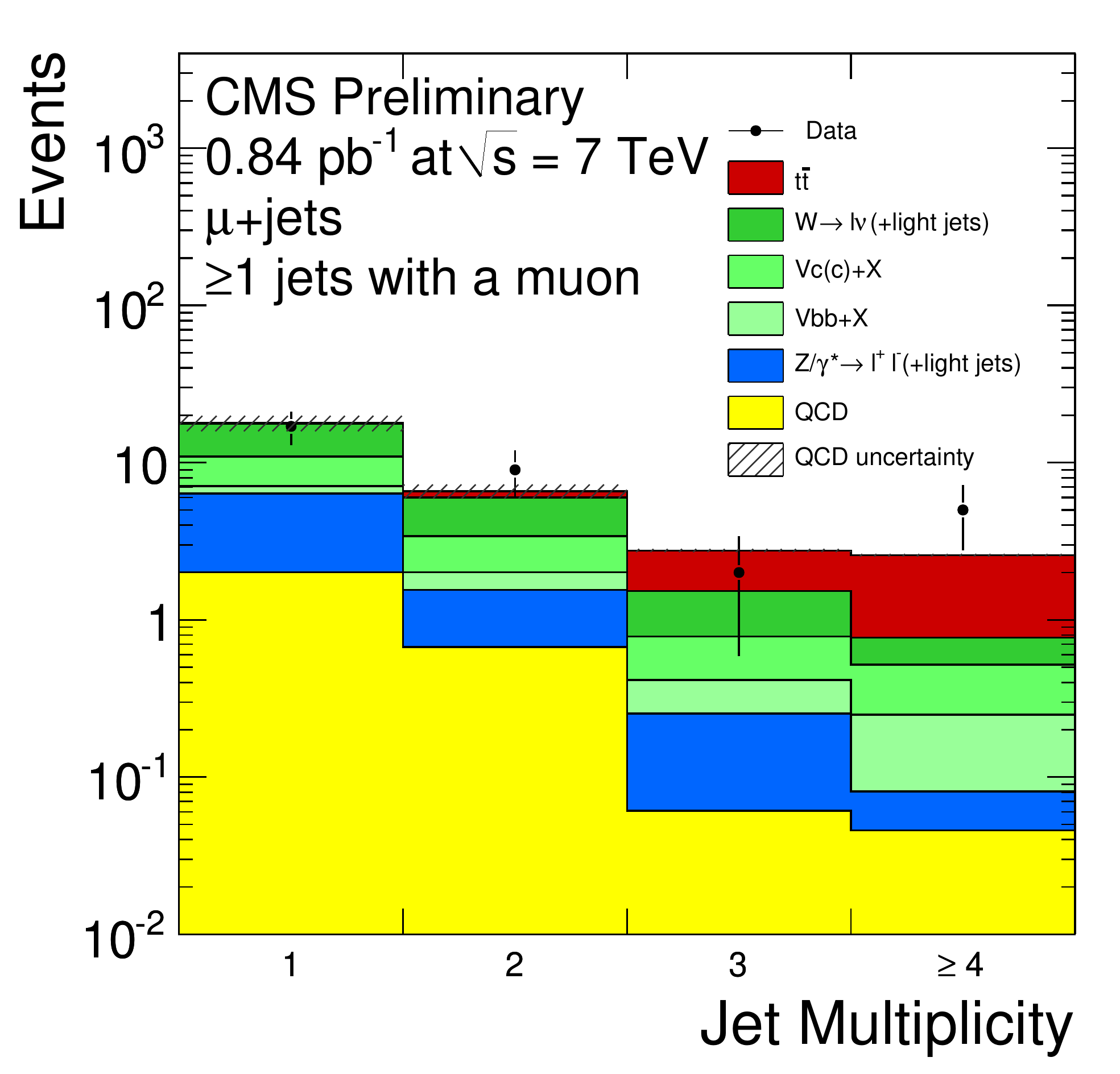}
\includegraphics[angle=0,width=0.45\linewidth]{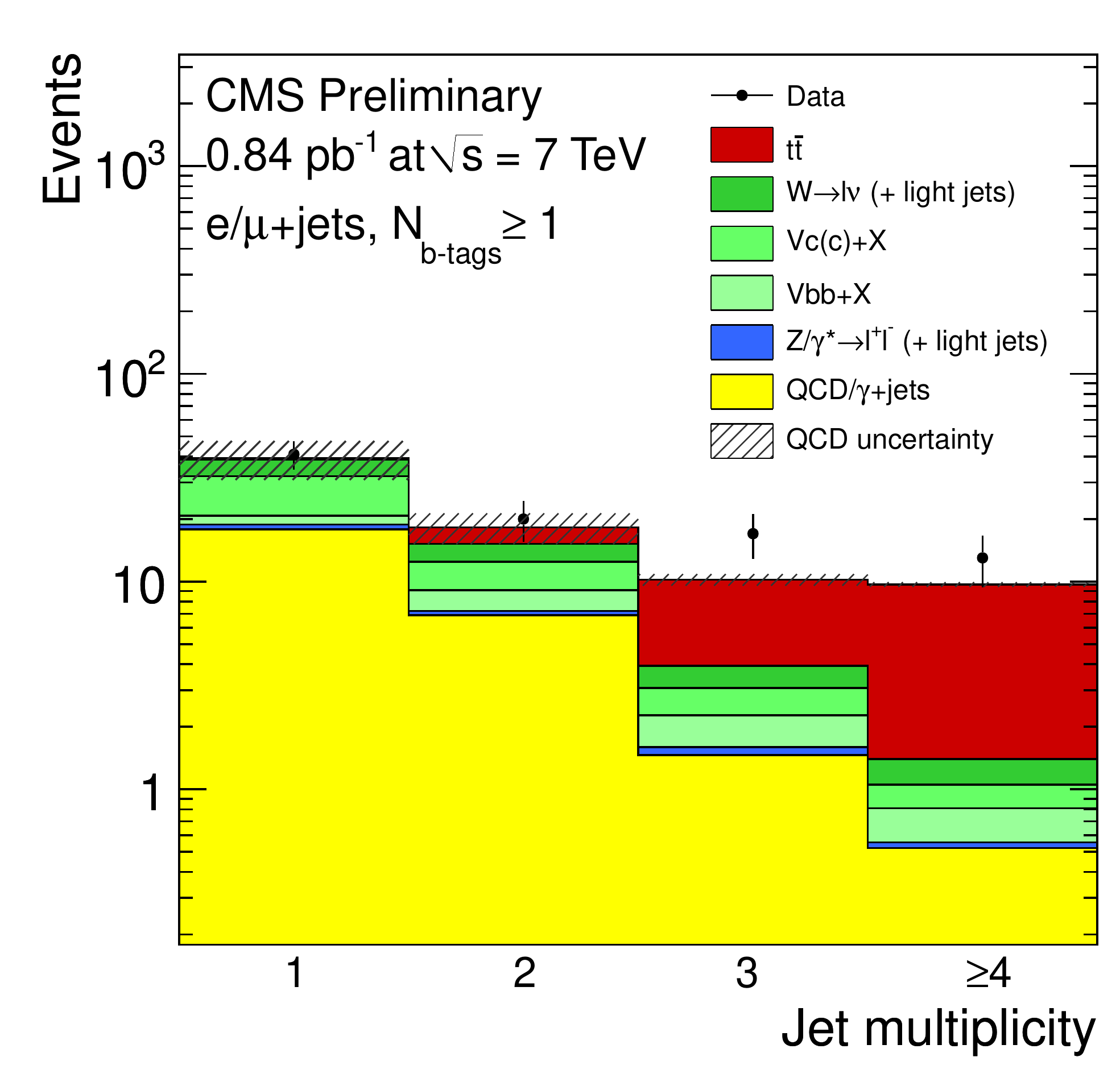}
\caption{Jet multiplicity for $\mu$+jets events, where one of the jets has an additional muon-in-jet signature (left), as well as for $e$+jets and $\mu$+jet combined, where at least one of the jets is b-tagged using a secondary vertex algorithm (right).}
\label{fig:ljets-btag}
\end{figure}

Since top quarks decay to b-quarks, one expects two b-jets to be present in every event in the case of signal, less so for the important background processes such as QCD and W+jets, which contain a mixture of light and heavy quarks. Therefore, the purity of the selection can be enhanced by selecting events containing b-jets. A simple way to enrich the b-content of the sample is by requesting the presence of at least one jet which contains a muon within $\Delta R<0.4$, typically originating from a semileptonic b-decay. The jet multiplicity for such a selection in $\mu$+jets is shown in Figure~\ref{fig:ljets-btag}. For $N_{\rm jets}\geq 3$, 7 events are observed in data, where 2.5 events are expected from non-top background. The W/Z+jets MADGRAPH samples contain a properly weighted mixture W/Z+light, W/Z+bb+jets and W/Z+c(c)+jets, where heavy quarks are produced from both gluon splitting as well as using heavy quark matrix elements, and combined using a matching procedure. Also shown in Figure~\ref{fig:ljets-btag} is the jet multiplicity for $e/\mu$+jets combined after requesting at least one b-tagged jet, using a secondary vertex tagger. For  $N_{\rm jets}\geq 3$, 30 events are observed in data, in a region where $\sim 5.5$ events are expected from non-top background. 


\section{CANDIDATE EVENTS}

\begin{figure}[tb]
\centering
\includegraphics[width=0.49\linewidth]{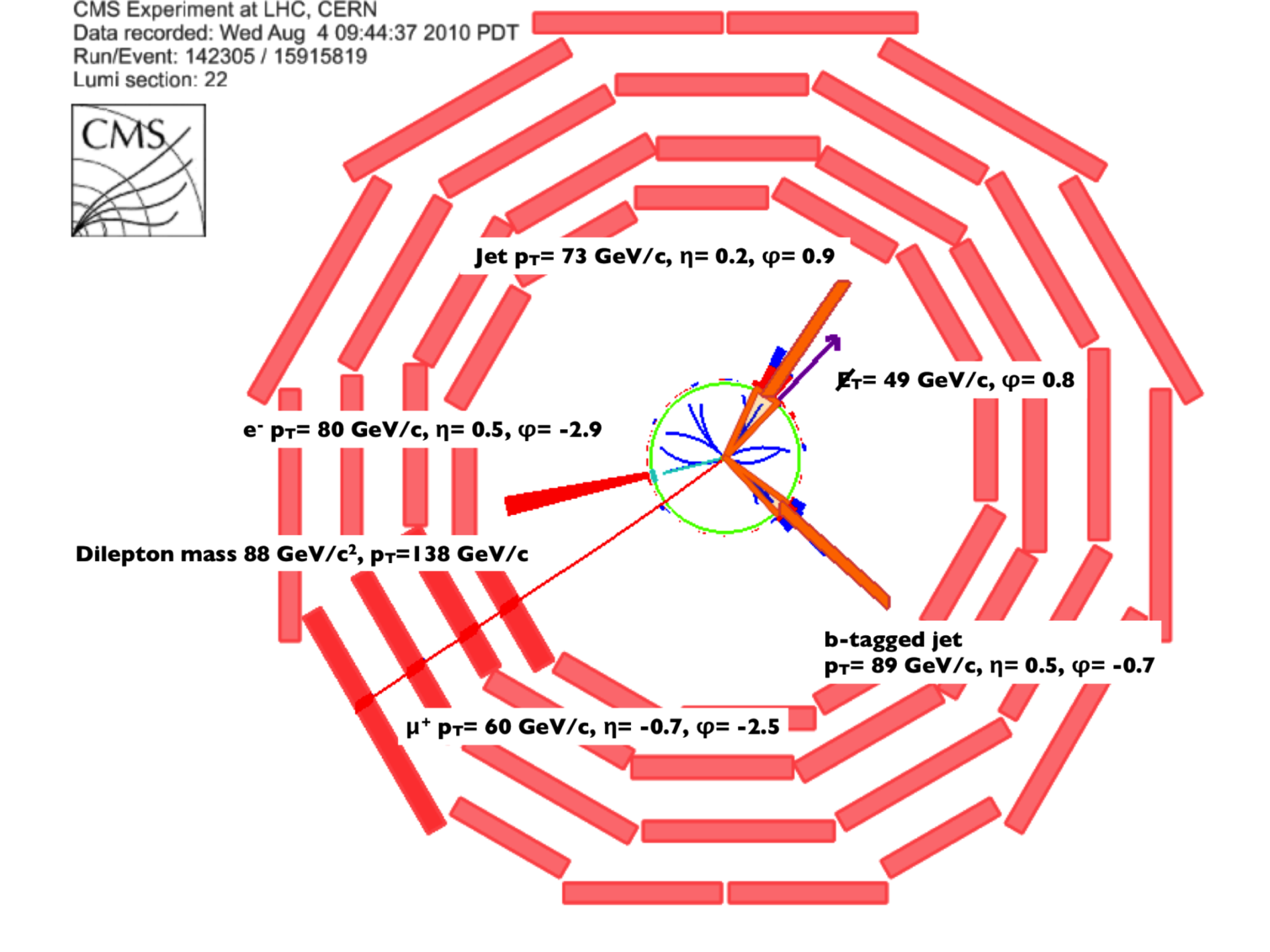}
\includegraphics[width=0.45\linewidth]{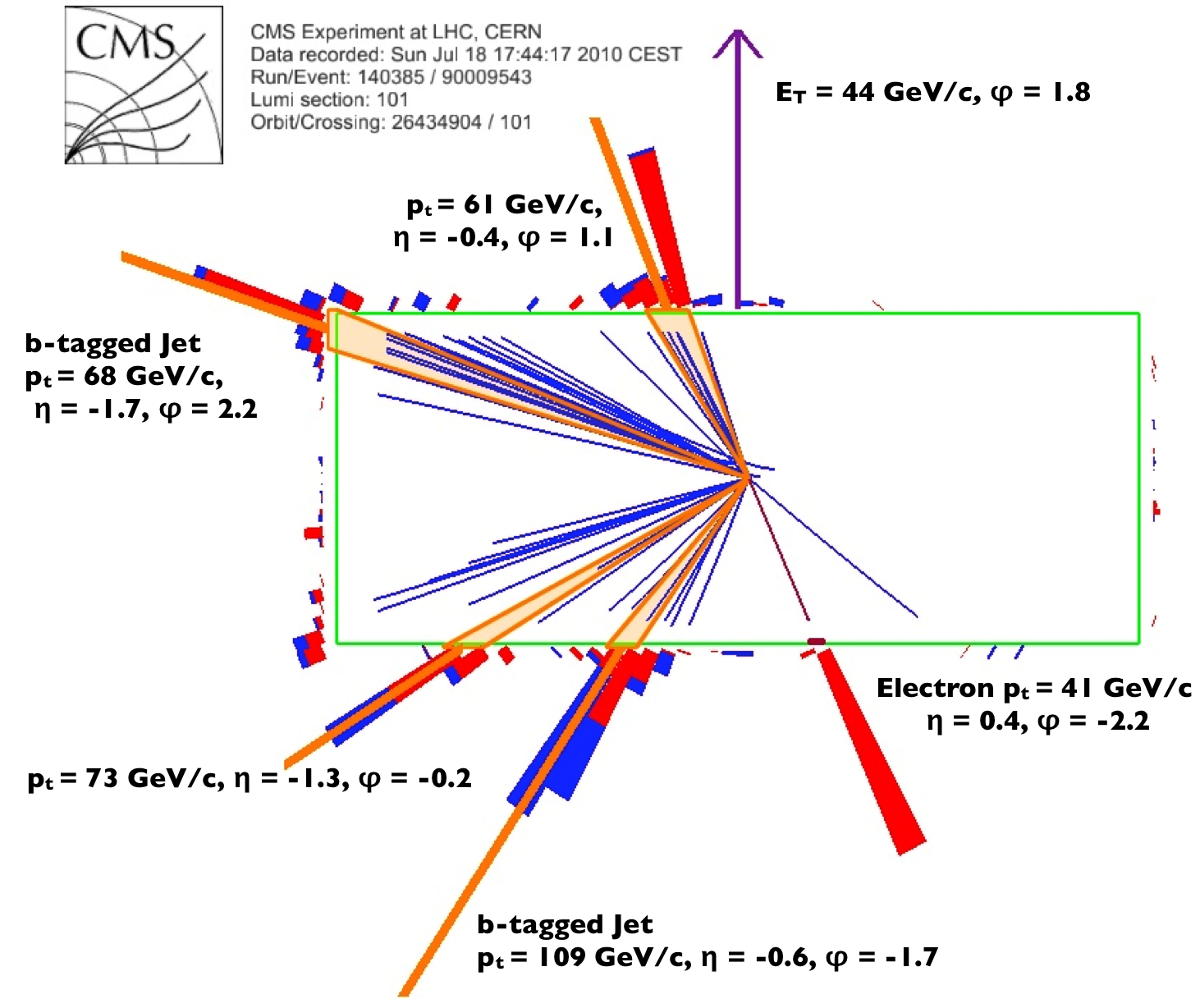}
\caption{Event displays of an $e\mu$ dilepton candidate in $r\phi$ view (left), and of an e+jets candidate in $\rho z$ view (right).}
\label{fig:evdisp}
\end{figure}

Two example candidate events are shown in Figure~\ref{fig:evdisp}. 
A dilepton candidate event is shown in Figure~\ref{fig:evdisp} (left). It has a muon with $p_T=60 \rm\ GeV/c$, an electron with $p_T=80 \rm\ GeV/c$, and large MET = 49 GeV. It contains two jets with $p_T=89$ and $73 \rm\ GeV/c$. One of the jets is b-tagged. The mass hypothesis is consistent with being a top event.

Figure~\ref{fig:evdisp} (right) shows a candidate event in the $e$+jets
mode passing the full event selection.  It has one isolated electron with $p_T=41 \rm\ GeV/c$, MET$=$44 GeV, and four high $p_T$ jets, with $p_T=$ 109, 73, 68
and $61 \rm\ GeV/c$, among which two are $b$-tagged.  The reconstructed
transverse W mass is $77 \rm\ GeV/c^2$, the invariant mass of the untagged
jets is $102 \rm\ GeV/c^2$, and the two possible hadronic top combinations,
the 3-jet system comprised of the two untagged jet and either the
highest or second highest $p_T$ tagged jets, have masses 232 and
$208 \rm\ GeV/c^2$, respectively.

\section{CONCLUSIONS}

In both the dilepton and the lepton+jets channel, events are observed in signal regions expected to be dominated by top quark pair production. The observed rates are roughly consistent with current theory expectations for top quark pair production, taking into account the experimental uncertainties due to e.g. jet energy scale, b-tagging performance, but also the theoretical uncertainties (e.g. scale and parton distributions for top signal, heavy flavour treatment for W/Z+jets backgrounds). The first top quark cross section measurements will come soon.


\begin{acknowledgments}

The author wishes to thank the members of the CMS Top Quark Physics Analysis Group for all their hard work which made it possible to produce the results presented in this document in a very short amount of time. 

\end{acknowledgments}


\end{document}